\documentclass[10pt]{iopart}
\usepackage{graphicx}
\usepackage{color}

\expandafter\let\csname equation*\endcsname\relax

\expandafter\let\csname endequation*\endcsname\relax

\usepackage[libertine]{newtxmath}
\usepackage{slashed}
\usepackage{amsfonts}
\usepackage[normalem]{ulem}

\usepackage{microtype}\usepackage[colorlinks,linkcolor=blue,anchorcolor=blue,citecolor=blue,urlcolor=blue,breaklinks=true]{hyperref}

\expandafter\let\csname equation*\endcsname\relax

\expandafter\let\csname endequation*\endcsname\relax

\def\be{\begin{equation}}
\def\ee{\end{equation}}
\def\bea{\begin{eqnarray}}
\def\eea{\end{eqnarray}}

\begin{document}

%\preprint{ADP-25-29/T1291}

\title{WIMP dark matter within the dark photon portal}

\author{Xuan-Gong Wang$^1$\footnote{Author to whom any correspondence should be addressed.}, B M Loizos$^1$, A W Thomas$^1$}
\address{$^1$ ARC Centre of Excellence for Dark Matter Particle Physics and CSSM, Department of Physics, University of Adelaide, Adelaide SA 5005, Australia}
\eads{\mailto{xuan-gong.wang@adelaide.edu.au}, \mailto{bill.loizos@adelaide.edu.au}, \mailto{anthony.thomas@adelaide.edu.au}}

\begin{abstract}
We test the dark photon as a portal connecting to the dark sector in the case of Dirac fermion and complex scalar dark matter with masses up to 1 TeV. Both the dark photon and the $Z$ boson contribute to the dark matter annihilation and dark matter--nucleon scattering processes. We derive the lower limits on the dark parameters from thermal relic density. The corresponding spin-independent dark matter--proton cross sections are compared with the upper bounds set by direct detection. We explore the allowed regions of the dark parameter space that are consistent with these constraints.
\end{abstract}


\submitto{\jpg}

\maketitle

%%%%%%%%%%%%%%%%%%%%%%%%%%%%%%%%%%%%%%%%%%%%%%%%%%%%%%
\section{Introduction}
Unveiling the nature of dark matter (DM) is a central goal in contemporary physics, yet its properties remain entirely unknown~\cite{Cirelli:2024ssz}. Among numerous well-motivated dark matter models, weakly interacting massive particles (WIMPs)~\cite{Bertone:2004pz, Jungman:1995df, Goodman:1984dc} are still amongst the most promising dark matter candidates, even though more effort is now being put into new DM scenarios~\cite{Battaglieri:2017aum}, such as light dark matter in sub-GeV region both theoretically~\cite{Batell:2014mga, Knapen:2017xzo, Gori:2025jzu} and experimentally~\cite{Banerjee:2019pds, COHERENT:2021pvd}.

Combined constraints from dark matter relic density, direct detection and indirect detection have been applied to test various dark matter models~\cite{Zheng:2010js, Yu:2011by, Balan:2024cmq}. The observed thermal relic density, $\Omega_{\rm DM} h^2 = 0.1200 \pm 0.0012$~\cite{ParticleDataGroup:2024cfk}, requires a sufficiently large cross section for dark matter annihilation to Standard Model (SM) particles to avoid overabundance, which could be applied to set lower limits on dark parameters such as coupling constants.
Direct detection, on the other hand, has placed stringent upper bounds on spin-independent (CRESST~\cite{CRESST:2019jnq, CRESST:2019axx}, DarkSide~\cite{DarkSide-50:2022qzh}, XENON~\cite{XENON:2018voc, XENON:2020gfr}, PandaX~\cite{PandaX-4T:2021bab}, and LZ~\cite{LZ:2022lsv, LZ:2024zvo}) and spin-dependent (LZ~\cite{LZ:2024zvo}, PandaX~\cite{PandaX-II:2016wea}, LUX~\cite{LUX:2017ree}, XENON~\cite{XENON:2019rxp}, and PICO~\cite{PICO:2019vsc}) cross sections, though subject to possible relativistic mean-field corrections~\cite{Wang:2020zgp} and  uncertainties from nuclear form factors~\cite{AbdelKhaleq:2023ipt}.

The dark photon model~\cite{Fabbrichesi:2020wbt, Filippi:2020kii} has received considerable attention, both in terms of its potential to provide new physics phenomena beyond the Standard Model (SM)~\cite{Graham:2021ggy, He:1991qd} and as a promising portal connecting to the dark sector~\cite{Boehm:2003hm, Hambye:2019dwd}. However, there are two versions of the dark photon model, with the dark photon kinetically mixing with either the Standard Model photon or the hypercharge $B$ boson, respectively,
\begin{equation}
{\cal L}_{\rm DP} = \left\{ \begin{array}{c}
\epsilon F'_{\mu\nu}F^{\mu\nu}\, ,\ ({\rm photon-mxing})\\
\ \\
\frac{\epsilon}{2\cos\theta_W} F'_{\mu\nu}B^{\mu\nu}\, ,\ ({\rm hypercharge-mxing})\, ,
\end{array}
\right.
\end{equation}
where $\epsilon$ is the mixing parameter and $\theta_W$ is the Weinberg angle.

In the photon-mixing model, the $Z$ boson mass and its couplings are not modified by the dark photon, $A'$. Therefore, the mixing parameter $\epsilon$ is less constrained by electroweak precision observables (EWPO)~\cite{Altarelli:1991fk,Degrassi:1996ps,Dubovyk:2019szj,Ciuchini:2013pca}. Moreover, only the dark photon $A'$ contributes to dark matter annihilation ($s$-channel) and to direct detection ($t$-channel) and there is one-to-one correspondence between these two processes. Stringent constraints have been placed on the dark parameters from thermal relic density and direct detection of dark matter in the sub-GeV region~\cite{Izaguirre:2015yja,Feng:2017drg, Krnjaic:2025noj}. 

The hypercharge-mixing model has richer phenomenological implications. As a result of the $A'-Z$ mixing, the dark parameters are strongly constrained by electroweak precision observables~\cite{Hook:2010tw, Curtin:2014cca}, followed by recent analyses~\cite{Loizos:2023xbj, Harigaya:2023uhg, Bento:2023flt, Davoudiasl:2023cnc}. It has been used to investigate the $W$ boson mass anomaly~\cite{Zhang:2022nnh, Zeng:2022lkk, Cheng:2022aau, Thomas:2022gib} associated with the CDF measurement~\cite{CDF:2022hxs}. Contrary to the photon-mixing model, in this case the dark photon also couples to neutrinos, contributing to rare kaon~\cite{Wang:2023css} and $B$ meson decays to neutrino states~\cite{Davoudiasl:2012ag, Datta:2022zng, Seto:2025mte}. Moreover, the dark photon has nonzero axial-vector couplings to SM fermions, therefore contributing to parity-violating electron scattering~\cite{Thomas:2022gib, Thomas:2022qhj}. 

In light of potential couplings of dark photons to dark matter particles, constraints on the mixing parameter from $e^+ e^-$~\cite{Abdullahi:2023tyk} and hadron colliders~\cite{Felix:2025afw} could be significantly relaxed. In particular, the $Z$ boson will also couple to dark matter particles~\cite{Loizos:2023xbj}. As a result, the dark matter annihilation and scattering processes will receive additional contributions from $Z$-boson exchange and $A'-Z$ interference. Only recently have these ideas been applied to thermal relic density and direct detection, focusing on Dirac fermion dark matter with masses in the GeV region~\cite{Alonso-Gonzalez:2025xqg} and the dark photon resonance region with $m_{\chi}\sim M_{A_D}/2$. 

In this work, we explore constraints on the parameter space of the dark sector within the hypercharge-mixing dark photon portal. Compared with Ref.~\cite{Alonso-Gonzalez:2025xqg}, (i) we extend the dark matter mass up to 1 TeV; (ii) we also present explicitly the resonance contribution from the $Z$ boson in the region of $m_{\rm DM} \approx M_Z/2$; (iii) we also investigate complex scalar dark matter in addition to the Dirac fermion scenario. 

We begin with a brief review of the dark photon formalism in Sec.~\ref{sec:dp-model}. We present the constraints from relic density and direct detection on the Dirac fermion and complex scalar dark matter in Sect.~\ref{sec:Dirac} and~\ref{sec:Scalar}, respectively. Finally, we present our conclusions in Sec.~\ref{sec:Conclusion}.

%%%%%%%%%%%%%%%%%%%%%%%%%%%%%%%%%%%%%%%%%%%%%%%%%%%%%%
\section{Dark photon formalism}
\label{sec:dp-model}

\subsection{Dark photon model}
The dark photon, $A'$, is usually introduced as an extra $U(1)$ gauge boson which kinetically mixes with the SM hypercharge $B$ boson~\cite{Fayet:1980ad, Fayet:1980rr, Holdom:1986eq, Okun:1982xi},
\begin{equation}
\label{eq:L}
{\cal L}_{\rm DP}  =  
- \frac{1}{4} F'_{\mu\nu} F'^{\mu\nu} + \frac{1}{2} m^2_{A'} A'_{\mu} A'^{\mu} 
+ \frac{\epsilon}{2 \cos\theta_W} F'_{\mu\nu} B^{\mu\nu}\, ,
\end{equation}
where $F'_{\mu\nu}$ is the field strength tensor, and $\theta_W$ is the Weinberg angle. The dark photon mass $m_{A'}$ could be generated from the dark Higgs~\cite{Galison:1983pa} or the Stueckelberg mechanism~\cite{Stueckelberg:1938hvi}, though the details do not affect the phenomenological analysis in this work.

By performing field redefinitions and diagonalizing the mass-squared matrix, the physical $Z$ and dark photon $A_D$ can be written in terms of the unmixed fields $\bar{Z}$ and $A'$,
\begin{align}
Z_{\mu} & = \cos{\alpha} \bar{Z}_{\mu} + \sin{\alpha} A'_{\mu}\, ,\nonumber\\
{A_{D}}_{\mu} & = -\sin{\alpha} \bar{Z}_{\mu} + \cos{\alpha} A'_{\mu}\, ,
\end{align}
where $\alpha$ is the $\bar{Z} - A'$ mixing angle~\cite{Kribs:2020vyk}
\begin{equation}
\tan{\alpha} = \frac{1}{2\epsilon_{W}} \bigg(1 - \epsilon_{W}^2 - \rho^2  - \text{sign}(1 - \rho^2) \sqrt{4 \epsilon_{W}^2 + (1 - \epsilon_{W}^2 - \rho^2)^2} \bigg)\ ,
\end{equation}
with
\begin{eqnarray}
\epsilon_W &=& \frac{\epsilon \tan{\theta_W}}{\sqrt{1 - \epsilon^2 / \cos^2\theta_W}}\, ,\nonumber\\
\rho &=& \frac{m_{A'}/m_{\bar{Z}}} {\sqrt{1 - \epsilon^2 / \cos^2\theta_W}}\, .
\end{eqnarray}
While the parameter $\epsilon$ in the photon-mixing model is unconstrained in prior, here the mixing parameter is bound by $\epsilon^{\rm max} = \cos\theta_W$. In addition, there is the so-called ``eigenmass repulsion" region in the $\epsilon-M_{A_D}$ plane associated with the physical masses~\cite{Kribs:2020vyk}, 
\begin{equation}
\label{eq:m_Z_AD}
M^2_{Z, A_D} = \frac{m_{\bar{Z}}^2}{2} [ 1 + \epsilon_W^2 + \rho^2 
 \pm {\rm sign}(1-\rho^2) \sqrt{(1 + \epsilon_W^2 + \rho^2)^2 - 4 \rho^2} ] \, ,
\end{equation}
in which the dark photon parameters are not accessible. This is because the mass difference is always finite for non-zero $\epsilon$, $|M^2_Z - M^2_{A_D}|\ge 2 |\epsilon| m^2_{\bar Z}$.

The couplings of the physical dark photon $A_D$ to SM fermions (in unit of $e = \sqrt{4\pi\alpha_{\rm em}}$) are given by~\cite{Kribs:2020vyk}
\begin{eqnarray}
\label{eq:C_AD}
C_{A_D}^v &=& - (\sin\alpha + \epsilon_W \cos\alpha) C_{\bar{Z}}^v + \epsilon_W \cos\alpha \cot \theta_W C_{\gamma}^v ,\nonumber\\
C_{A_D}^a &=& - (\sin\alpha + \epsilon_W \cos\alpha) C_{\bar{Z}}^a 
\, .
\end{eqnarray}
The Standard Model couplings of the $Z$ boson will be shifted to the physical ones,
\bea
\label{eq:C_Z}
C_{Z}^v &=& (\cos\alpha - \epsilon_W \sin\alpha) C_{\bar{Z}}^v + 
\epsilon_W \sin\alpha \cot \theta_W C_{\gamma}^v ,\nonumber\\
C_{Z}^a &=& (\cos\alpha - \epsilon_W \sin\alpha) C_{\bar{Z}}^a\, ,
\eea 
where $C_{\gamma}^v$ are the electromagnetic couplings,
\be
\{ C^v_{\gamma,\nu}, C^v_{\gamma,e}, C^v_{\gamma,u}, C^v_{\gamma,d}\} = \{ 0, -1, 2/3, - 1/3 \}\, ,
\ee
and $C_{\bar{Z}}^{v(a)}$ are the vector (axial-vector) SM weak couplings~\cite{Wang:2023css, Kribs:2020vyk},
\begin{align}
\big\{
C^v_{\bar{Z},\nu}, C^v_{\bar{Z},e}, C^v_{\bar{Z},u}, C^v_{\bar{Z},d} 
\big\}\, \sin 2\theta_W
&=
\Big\{ 
\frac{1}{2}, - \frac{1}{2} + 2 \sin^2\theta_W,\,
  \frac{1}{2} - \frac{4}{3}\sin^2\theta_W,\,
- \frac{1}{2} + \frac{2}{3}\sin^2\theta_W 
\Big\}\, ,\nonumber\\
\big\{
C^a_{\bar{Z},\nu}, C^a_{\bar{Z},e}, C^a_{\bar{Z},u}, C^a_{\bar{Z},d} 
\big\}\, \sin 2\theta_W
&= \Big\{ \frac{1}{2}, - \frac12,\, \frac12,\, - \frac12 \Big\}\, . 
\end{align}
Using these couplings, the decay width of dark photon to SM final state is
\be
\Gamma_{A_D \to {\rm SM}} =  \sum_{f} N_C^f \cdot \frac{M_{A_D} \alpha_{\rm em}}{3} \bigg\{\left(1 + \frac{2m_{f}^2}{M_{A_D}^2}\right) (C^v_{A_D,f})^2 
+ \left(1 - \frac{4m_{f}^2}{M_{A_D}^2}\right) (C^a_{A_D,f})^2\bigg\}\sqrt{1 - \frac{4m_{f}^2}{M_{A_D}^2}}\, ,
\ee
where $N_C^f = 1$ for leptons and $N_C^f = 3$ for quarks.

The couplings to nucleons can be derived as linear combinations of the couplings to the $u$ and $d$ quarks,
\bea
\label{eq:C_pn}
C_{A_D(Z),p}^{v,a} &=& 2 C_{A_D (Z),u}^{v,a} + C_{A_D (Z),d}^{v,a} \, ,\nonumber\\
C_{A_D(Z),n}^{v,a} &=& C_{A_D(Z),u}^{v,a} + 2 C_{A_D(Z),d}^{v,a}\, .
\eea

\subsection{Dark matter model}
\label{sec:dm-model}
The dark photon is also a promising portal connecting to dark matter particles. Popular scenarios include Dirac, pseudo-Dirac, scalar and asymmetric dark matter models~\cite{Izaguirre:2015yja,Feng:2017drg, Krnjaic:2025noj}. In this work, we will focus on Dirac fermion $\chi$ and complex scalar $\phi$ which interact with the dark photon through
\bea
{\cal L}_{\chi} & = & \bar{\chi} ( i \gamma^{\mu} \partial_{\mu} - m_{\chi} ) \chi + g_{\chi} \bar{\chi}\gamma^{\mu} \chi A'_{\mu}\, ,\nonumber\\
{\cal L}_{\phi} & = & \partial_{\mu}\phi^{*} \partial^{\mu}\phi - m^2_{\phi} \phi^{*} \phi + i g_{\phi} (\phi^{*} \partial^{\mu} \phi - \phi \partial^{\mu} \phi^{*} ) A'_{\mu}\, .
\eea
The physical couplings can be written as
\be
\label{eq:C-AD-DM}
C^v_{A_D,\bar{\chi}\chi (\phi^* \phi)} = \frac{g_{\chi (\phi)} \cos\alpha}{\sqrt{1 - \epsilon^2/\cos\theta^2_W}}\, .
\ee
Here, $g_{\chi(\phi)}$ is typically of ${\cal O}(1)$ which may lead to a very large decay width, especially in the case of Dirac fermion dark matter,
\bea
\label{eq:Gamma-AD}
\Gamma_{A_D \to \bar{\chi}\chi} &=& \frac{M_{A_D} ( C^v_{A_D,\bar{\chi}\chi})^2}{12\pi} \left(1 + \frac{2m_{\chi}^2}{M_{A_D}^2}\right) \sqrt{1 - \frac{4m_{\chi}^2}{M_{A_D}^2}}\, ,\nonumber\\
\Gamma_{A_D \to \phi^{*}\phi} &=&\frac{M_{A_D} (C^v_{A_D,\phi^{*}\phi})^2}{48\pi} \left(1 - \frac{4 m_{\phi}^2}{M_{A_D}^2}\right) \sqrt{1 - \frac{4m_{\phi}^2}{M_{A_D}^2}}\, .
\eea
Notably, the $Z$ boson will also couple to dark matter particles with the physical coupling 
\be
C^v_{Z,\bar{\chi}\chi (\phi^* \phi)} = \frac{g_{\chi (\phi)} \sin\alpha}{\sqrt{1 - \epsilon^2/\cos\theta^2_W}}\, ,
\ee
therefore contributing to both the dark matter annihilation and DM-nucleon scattering processes.

\subsection{Thermal relic}
\label{sec:thermal-relic}
In the thermal freeze out framework, the DM number density $n_{\rm DM}$ evolves as~\cite{Cirelli:2024ssz}
\be
\dot{n}_{\rm DM} + 3 H n_{\rm DM} = \langle \sigma v \rangle [(n^{\rm eq}_{\rm DM})^2 - n^2_{\rm DM}]\, ,
\ee
where $H$ is the Hubble rate, $n^{\rm eq}_{\rm DM}$ is the number density that DM particles would have in thermal equilibrium and $\langle \sigma v \rangle$ is the thermally-averaged annihilation cross section~\cite{Gondolo:1990dk, Griest:1990kh}. 

Away from the $Z$ pole, the dark matter annihilation cross section, $\sigma({\rm DM} + {\rm DM} \to \bar{f} + f)$, is dominated by the dark photon exchange~\cite{Alonso-Gonzalez:2025xqg}. In the limit $M_{A_D} \gg m_{\rm DM}, \Gamma_{A_D}$, $\langle \sigma v \rangle$ depends on the dimensionless combination~\cite{Filippi:2020kii, Izaguirre:2015yja}
\be
\label{eq:y}
y = \epsilon^2 \alpha_D \left( \frac{m_{\rm DM}}{M_{A_D}} \right)^4\, ,
\ee
where $\alpha_D = g^2_{\chi (\phi)}/4\pi$. For a given value of $m_{\rm DM}$, the dark matter relic density and the spin-independent DM-nucleon scattering cross sections are insensitive to separate factors, $\epsilon$, $\alpha_D$ and the mass ratio $R=M_{A_D}/m_{\rm DM}$.

In searches over a wider range of parameter space, the annihilation cross section will exhibit strong sensitivities to individual factors $R$ and $\alpha_D$. In particular, as emphasized in Ref.~\cite{Feng:2017drg}, in the resonance region where $R \approx 2$ there is a very strong increase in the annihilation cross section. Also, the $Z$ boson contributions will be significant in the region $2 m_{\rm DM} \approx M_Z$.

\subsection{Direct detection}

The upper bounds on the spin-independent DM-nucleon cross section set by direct detection assume $\sigma_{{\rm DM}, p} = \sigma_{{\rm DM}, n}$, so that the event rate is proportional to $A^2 \cdot \sigma^{\rm SI}_{{\rm DM}, p}$. However, in the present model, the leading contributions to the DM-nucleus scattering in the non-relativistic limit come from the standard spin-independent interaction, ${\cal O}_1 = \mathbf{1}_{\chi} \mathbf{1}_N$, with the couplings being 
\bea
\label{eq:c1}
c_1^p &=& \frac{C^v_{A_D,\bar{\chi}\chi} \cdot e C^v_{A_D,p}}{M^2_{A_D}} + \frac{C^v_{Z,\bar{\chi}\chi} \cdot e C^v_{Z,p}}{M^2_{Z}} \, ,\nonumber\\
c_1^n &=& \frac{C^v_{A_D,\bar{\chi}\chi} \cdot e C^v_{A_D,n}}{M^2_{A_D}} + \frac{C^v_{Z,\bar{\chi}\chi} \cdot e C^v_{Z,n}}{M^2_{Z}} \, .
\eea
Substituting Eqs.~(\ref{eq:C_pn}) and (\ref{eq:C_AD}-\ref{eq:C_Z}) into Eq.~(\ref{eq:c1}), the coupling to the neutron is
\be
\label{eq:CAA}
c_1^{n} = \frac{g_{\chi} e (C^v_{\bar{Z}, u} + 2 C^v_{\bar{Z},d})}{\sqrt{1 - \epsilon^2/\cos^2\theta_W}} 
\cdot 
\left\{ - \frac{\cos\alpha (\sin\alpha + \epsilon_W \cos\alpha)}{M^2_{A_D}} 
+ \frac{\sin\alpha (\cos\alpha - \epsilon_W \sin\alpha)}{M^2_{Z}} \right\}\, ,
\ee
in which the electromagnetic components have been canceled because the neutron is electrically neutral, i.e., $C^v_{\gamma,u} + 2 C^v_{\gamma,d} = 0$.
In the limit $\epsilon \ll 1$, we can get~\cite{Alonso-Gonzalez:2025xqg}~\footnote{The mixing parameter $\epsilon$ in Ref.~\cite{Alonso-Gonzalez:2025xqg} is equivalent to $- \epsilon/\cos\theta_W$ in our work.}
\bea
\sin\alpha &=& - \eta \frac{\epsilon}{\cos\theta_W} + {\cal O}(\epsilon^3)\, ,\nonumber\\
\cos\alpha &=& 1 - \frac{\eta^2}{2} \frac{\epsilon^2}{\cos\theta^2_W}\, ,
\eea
where $\eta = \sin\theta_W/(1 -r)$ with $r = M^2_{A_D}/M^2_Z$. Therefore, the terms in the bracket of Eq.~(\ref{eq:CAA}) become
\be
\frac{1}{M^2_{A_D}} \cdot \left\{ - ( \sin\alpha + \epsilon_W ) + \sin\alpha \cdot r \right\}
= \frac{1}{M^2_{A_D}} \cdot \{ \epsilon \tan\theta_W - \epsilon_W \}\, ,
\ee
which exactly cancel, leading to a vanishing coupling of $c^n_1$ at this order.
As a result, the direct detection event rate scales as $Z^2 \cdot \sigma^{\rm SI}_{{\rm DM}, p}$. Therefore, as emphasized by Alonzo-Gonz\'alez {\em et al.}, these upper limits of direct detection should be relaxed~\cite{Alonso-Gonzalez:2025xqg}, by roughly a factor of $A^2/Z^2$, as given in Tab~\ref{table:rescaling-DD}.
\begin{table*}[!t]
\renewcommand\arraystretch{1.5}
\begin{center}
\begin{tabular}{cccc} \hline\hline
 Direct detection   &   Nuclear target    &   $(Z,A)$   &   $A^2/Z^2$ \\ \hline
 CRESST III (${\rm Ca W O_4}$)        &   O when $m_{\rm DM} < 5\ {\rm GeV}$          &    $(8,16)$     &   $4.00$ \\
  \                                                          &   W when $m_{\rm DM} > 5\ {\rm GeV}$         &    $(74,182)$  & $6.05$  \\
 DarkSide-50        &        Ar           &    $(18,40)$    &    $4.94$ \\
 XENON1T, PandaX, LZ  &       Xe          &    $(54, 131)$     &   $5.89$ \\ \hline\hline
\end{tabular}
\caption{The re-scaling factors for direct detection constraints. }
\label{table:rescaling-DD}
\end{center}
\end{table*}

\subsection{EWPO and collider constraints}
With the inclusion of a dark photon, a bunch of observables in the electroweak sector will be modified because of shifts in the $Z$ boson mass and its weak couplings. Strong constraints have been placed on the mixing parameter by fitting electroweak precision observables (EWPO) measured at lepton (LEP, SLC) and hadron (Tevatron, LHC) colliders~\cite{ParticleDataGroup:2024cfk}, with upper limits being of ${\cal O}(10^{-2})$~\cite{Hook:2010tw, Curtin:2014cca}. 
In the presence of dark matter particles, the $Z$ boson decay width will receive additional contributions from $Z \to \bar{\chi} \chi (\phi^* \phi)$ channels, and the constraints on $\epsilon$ are slightly relaxed~\cite{Loizos:2023xbj}.

Direct searches for dark photons at $e^+ e^-$~\cite{BaBar:2014zli} and hadron~\cite{LHCb:2019vmc, CMS:2019buh} colliders have led to more stringent constraints on the mixing parameter, $\epsilon \le 10^{-3}$, under the assumption that the dark photon only decays to SM final states. These can be significantly relaxed if the dark photon has a larger decay width in light of couplings to dark matter particles~\cite{Abdullahi:2023tyk, Felix:2025afw, Alonso-Gonzalez:2025xqg, CMS:2024zqs}. In the present work, we do not show those constraints as they are comparable with the EWPO limits, either from an analysis of cross sections~\cite{Felix:2025afw} or from a simple re-scaling by decay widths~\cite{Alonso-Gonzalez:2025xqg}.

In this work, we extend the EWPO constraints on $\epsilon$~\cite{Curtin:2014cca, Loizos:2023xbj} to $M_{A_D} = 3\ {\rm TeV}$, which are converted to upper limits on $y$ in the $y-m_{\rm DM}$ plane.

%%%%%%%%%%%%%%%%%%%%%%%%%%%%%%%%%%%%%%%%%%%%%%%%%%%%%
\section{Results}
We consider the dark matter mass $m_{\rm DM}$ up to 1 TeV. The dark photon mass $M_{A_D}$ then lies in the range up to a few TeV, depending on the ratio $R= M_{A_D}/m_{\rm DM}$.

Previous analyses~\cite{Izaguirre:2015yja, Feng:2017drg, Krnjaic:2025noj} usually start with the mass ratio $R=3$. Qualitatively, the resonance region with $R \approx 2$ is favoured by the existing constraints. The dark coupling can vary from the perturbativity bound of $\alpha_D =1/2$ down to $\alpha_{\rm em}$ or even lower. Following these earlier analyses, we take $R=3,\ 2.3,\ 2.05$ and $\alpha_D = 0.5,\ 0.05,\ 0.005$, which sufficiently cover the interesting region of the parameter space.

In our numerical analysis, we implemented the dark photon model into FeynRules~\cite{Christensen:2008py, Alloul:2013bka} and micrOMEGAs~\cite{Alguero:2023zol}. 
For given values of $R$ and $\alpha_D$, we adjust the mixing parameter $\epsilon$ to generate the observed dark matter relic density~\cite{ParticleDataGroup:2024cfk} 
\be
\Omega = \Omega_{\rm DM} h^2 = 0.1200 \pm 0.0012\, .
\ee
This sets the lower limit on $\epsilon$ to avoid overabundance, which can be converted to a lower bound on the variable $y$. Meanwhile, it also places a lower bound on the spin-independent (SI) DM-nucleon scattering cross section.
As noted earlier, the numerical values of the SI DM-neutron cross sections evaluated using Eq.~(\ref{eq:c1}) are negligible compared with those of DM-proton scattering.

\subsection{Dirac fermion}
\label{sec:Dirac}
\begin{figure}[!htbp]
\begin{center}
\includegraphics[width=\textwidth]{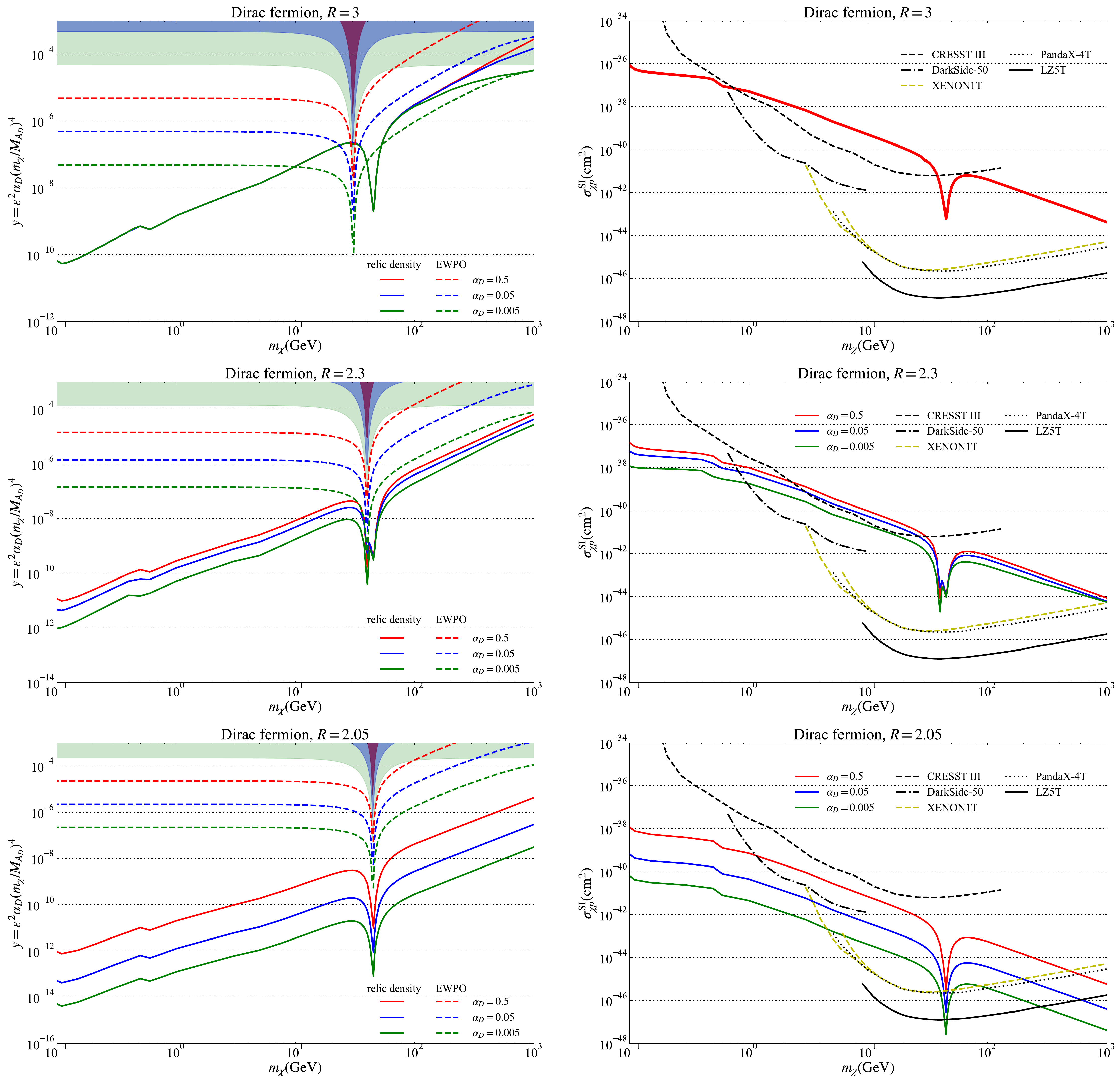}
\vspace*{-0.2cm}
\caption{(Left panels): Lower limits on $y$ from the dark matter relic density (solid lines) and the EWPO constraints (dashed lines). The latter  are derived by converting the exclusion limits on $\epsilon$ from Ref.~\cite{Curtin:2014cca, Loizos:2023xbj} with $M_{A_D}$ being extended up to 3 TeV. The shaded areas are the eigenmass repulsion regions~\cite{Kribs:2020vyk}, corresponding to different values of $\alpha_D$, in which the dark photon parameters are not accessible. (Right panels): The corresponding lower limits on $\sigma_{\chi p}^{\rm SI}$. We also show the exclusion constraints from CRESST III~\cite{CRESST:2019jnq, CRESST:2019axx}, DarkSide-50~\cite{DarkSide-50:2022qzh}, XENON1T~\cite{XENON:2018voc,XENON:2020gfr}, PandaX-4T~\cite{PandaX-4T:2021bab} and LZ5T~\cite{LZ:2024zvo}.}
\label{fig:Dirac-fermion}
\end{center}
\end{figure}

The lower limits on the variable $y$ from thermal relic density are shown in the left panels of Fig.~\ref{fig:Dirac-fermion}. The corresponding lower bounds on the SI DM-proton cross sections are given in the right panels, which are compared with the upper limits set by direct detection after taking into account the re-scaling factors in Tab.~\ref{table:rescaling-DD}.   

In the case $R=3$, the dark photon mass is far away from the resonant region. Therefore, its decay width plays a less important role, and the constraints on $y$ are not sensitive to $\alpha_D$. For $m_{\chi} < 100\ {\rm GeV}$, the lower limits of $y$ with $\alpha_D = 0.5,\ 0.05$ and $0.005$ coincide. In the high-mass region, there are discrepancies among these three cases because the physical couplings in Eq.~(\ref{eq:C-AD-DM}) deviate from $g_{\chi}$ due to large values of $\epsilon$. A new feature is that, in the region of $2 m_{\chi} \approx M_Z$, the resonant contribution from the $Z$ boson will be significant. The required couplings and the corresponding SI DM-proton cross sections could drop by as much as two orders of magnitude. In addition, we found that $\alpha_D$ cannot be arbitrarily small, otherwise the lower bounds of $y$ will exceed the EWPO constraints; as illustrated for example by the green lines with $\alpha_D = 0.005$. In all cases, Dirac fermion dark matter with masses in the GeV--TeV range is ruled out by direct detection.  

When the dark photon mass is close to the dark matter threshold, for example $R=2.3$, the dark matter annihilation cross section will be enhanced from the dark photon propagator. Smaller values of $y$ are required, which escape the EWPO limits for $\alpha_D = 0.5,\ 0.05$ and $0.005$.
However, the SI DM-proton cross sections still lie above the direct detection constraints for $M_{A_D}$ above about 2 GeV (for $\alpha_D=0.005$).

In the resonant annihilation regime with $R=2.05$, the lower limits of $y$ decrease further and the effect of the dark photon decay width is more significant. With a Breit-Wigner parametrisation, the dark photon contribution to the $s$-channel annihilation cross section~\cite{Alonso-Gonzalez:2025xqg}
\be
\label{eq:sigma-BW}
\sigma \propto \frac{\epsilon^2 \alpha_{\rm em} \alpha_D}{(s-M^2_{A_D})^2 + M^2_{A_D} \Gamma^2_{A_D}}\, .
\ee
In the case of $\alpha_D = 1/2$, the dark photon decay width is dominated by the $\bar{\chi}\chi$ channel in Eq.~(\ref{eq:Gamma-AD}), $\Gamma_{A_D} \approx \Gamma_{A_D\to \bar{\chi}\chi} \approx 0.05\ M_{A_D}$, which is non-negligible as the two terms in the denominator of Eq.~(\ref{eq:sigma-BW}) are compatible,
\be
s-M^2_{A_D} \approx 4 m^2_{\chi} - M^2_{A_D} = -0.05 M^2_{A_D}\, ,\ \ \ M_{A_D} \Gamma_{A_D} \approx 0.05 M^2_{A_D}\, .
\ee
For $\alpha_D = 0.05$ and $0.005$, the decay widths $\Gamma_{A_D}$ are much smaller and may therefore be neglected.
As shown in the lower panels of Fig.~\ref{fig:Dirac-fermion}, there is a strong dependence not just on $y$ but also on the separate factor $\alpha_D$. In this case there are indeed a few regions in which the SI DM-proton cross sections do not exceed the direct detection bounds.

Our results concerning Dirac fermion dark matter in the GeV region are qualitatively consistent with those reported recently in Ref.~\cite{Alonso-Gonzalez:2025xqg}. 
Moreover, the dark parameters are also constrained by indirect detection of dark matter, such as gamma rays from dwarf spheroidal galaxies~\cite{McDaniel:2023bju} and the Cosmic Microwave Background (CMB)~\cite{Planck:2018vyg}. A discussion of this for dark fermions may be found in Ref.~\cite{Alonso-Gonzalez:2025xqg}. In the next section, we will investigate complex scalar dark matter, which is one of representative DM scenarios that are safe from the CMB limits~\cite{Izaguirre:2015yja, Krnjaic:2025noj}.

\subsection{Complex scalar}
\label{sec:Scalar}

\begin{figure*}[!t]
\begin{center}
\includegraphics[width=\textwidth]{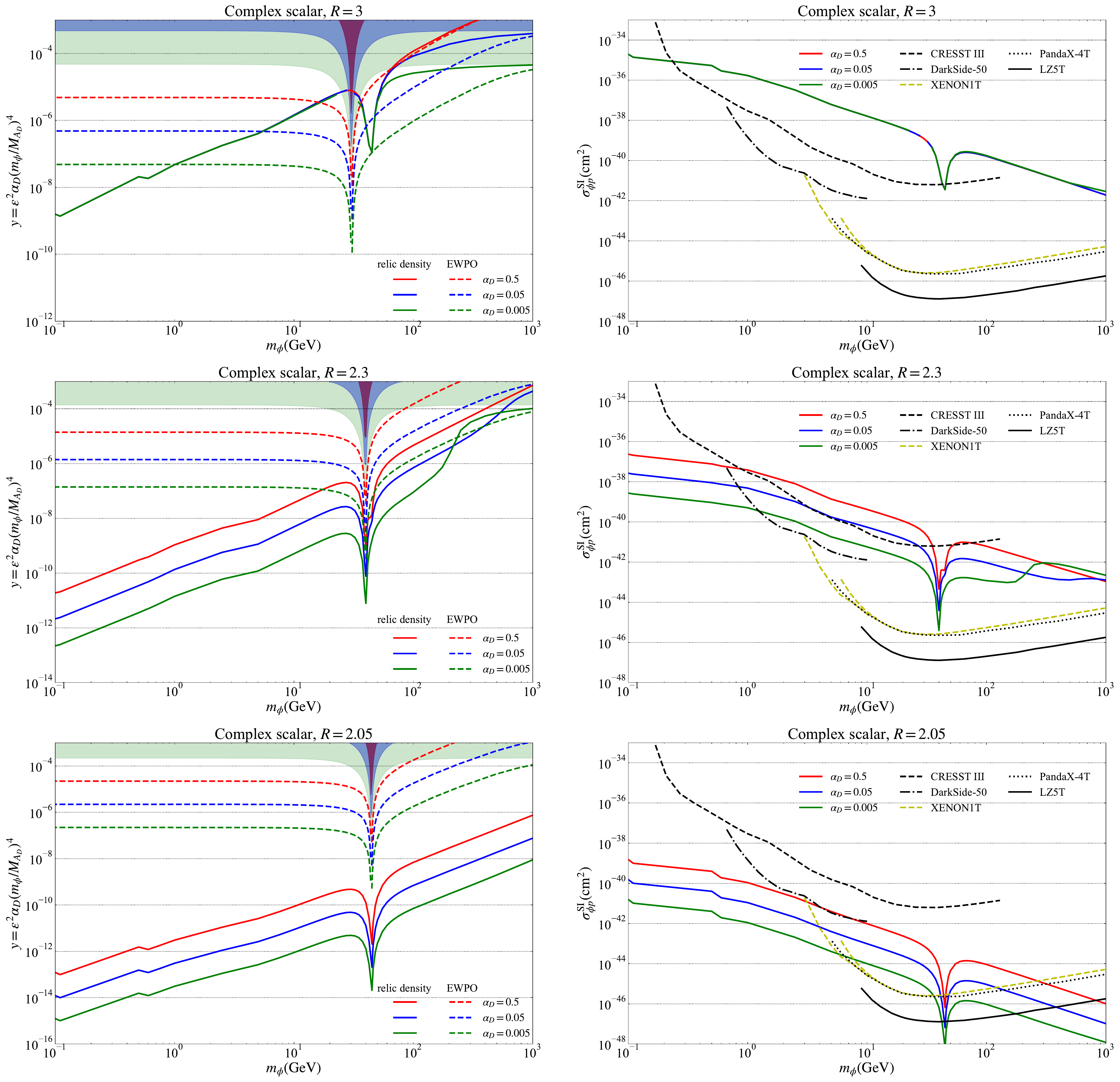}
\vspace*{-0.2cm}
\caption{The same as Fig.~\ref{fig:Dirac-fermion} but for the case of complex scalar dark matter.}
\label{fig:scalar}
\end{center}
\end{figure*}

In the case of complex scalar DM, we also consider $\alpha_D = 0.5,\ 0.05$ and $0.005$. The lower limits on $y$ and the corresponding SI DM-proton scattering cross sections are given in Fig.~\ref{fig:scalar}.

For $R=3$, the $s$-channel dark matter annihilation cross section will be suppressed by ${\cal O}(v^2)$, where $v$ is the dark matter thermal velocity at freeze out, which is typically of $v \sim c/4$~\cite{Gondolo:1990dk}. 
As a result, the required mixing parameter, or equivalently the variable $y$, needs to be larger than for Dirac fermionic DM. In most regions of the parameter space allowed by the EWPO constraints, the dark matter density will be overabundant. In particular, for $\alpha_D=0.05$ and $0.005$, there are no solutions for $y$ because of the eigenmass repulsion when $m_{\phi}$ lies in the range $[28.4, 32.2]\ {\rm GeV}$ and $[24.4, 34.8]\ {\rm GeV}$, respectively. Similarly to the Dirac fermion case, the SI DM-proton cross sections are much larger than the upper limits set by direct detection.

Interestingly, the variable $y$ and the SI DM-proton cross sections have a much stronger dependence on $R$ compared with that found for Dirac fermion dark matter. Reducing $R$ from 3 to 2.05, both $y$ and $\sigma^{\rm SI}_{\phi p}$ are found to decrease by four ($\alpha_D=0.5$, red solid lines) to six ($\alpha_D=0.005$, green solid lines) orders of magnitude. For $\alpha_D=0.005$ there is a relatively large region,  ($m_{\phi}< 6\ {\rm GeV}$ or $m_{\phi} > 180\ {\rm GeV}$, as well as a narrow region of $2 m_{\phi} \approx M_Z$), in which the predictions are consistent with all constraints from EWPO, thermal relic density, and direct detection. For the other two cases, $\alpha_D=0.5$ and $\alpha_D=0.05$, the region in which the results are compatible with all constraints is much narrower.

\begin{figure*}[!t]
\begin{center}
\includegraphics[width=\textwidth]{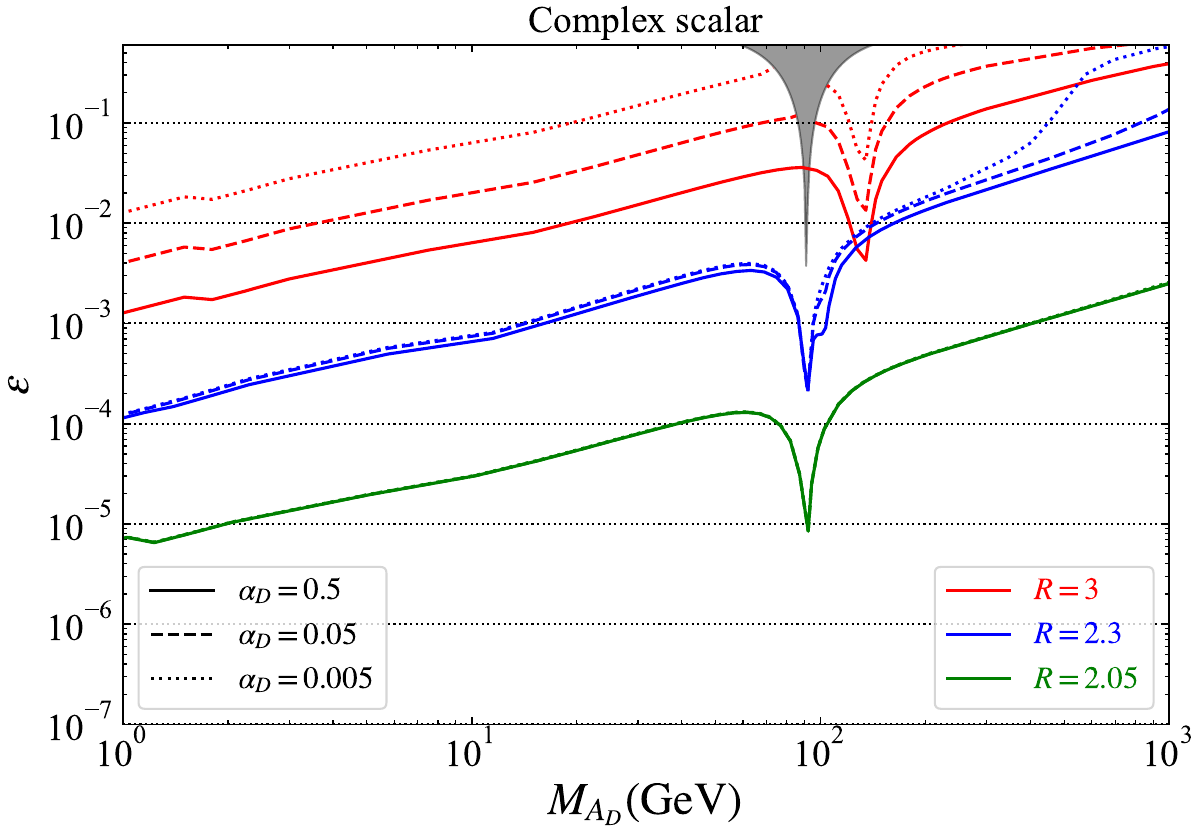}
\vspace*{-0.2cm}
\caption{The mixing parameter $\epsilon$ required by the thermal relic density constraint, in the case of complex scalar dark matter. Regions below these limits (smaller values of $\epsilon$) will lead to overabundance, and therefore are excluded.}
\label{fig:relic_epsilon_scalar}
\end{center}
\end{figure*}

In Fig.~\ref{fig:relic_epsilon_scalar}, we also show the separate parameter $\epsilon$ required by the relic density for each case. In some region of the parameter space, there may be cancellation among contributions from the dark photon, $Z$ boson, and their interference, suppressing the dark matter annihilation cross section. As a result, larger values of $\epsilon$ are required, e.g., in the heavy mass region with $R=2.3$ and $\alpha_D = 0.005$ (blue dotted line in Fig.~\ref{fig:relic_epsilon_scalar}), leading to intersections in the variable $y$ in Fig.~\ref{fig:scalar}.

%

%=======================================================%

\section{Conclusion}
\label{sec:Conclusion}

We have investigated the dark photon as a potential portal connecting the Standard Model to the dark sector, considering scenarios where dark matter consists of either a Dirac fermion or a complex scalar, with masses up to 1 TeV. In the hypercharge-mixing model, both the dark photon and the $Z$ boson contribute to dark matter annihilation, as well as to dark matter–nucleon scattering processes. We placed lower limits on the dimensionless variable $y$ from the thermal relic density, using typical values of the mass ratio $R= M_{A_D}/m_{\rm DM}$ and the separate factor $\alpha_D$. 

Both Dirac fermion and complex scalar dark matter are ruled out in the GeV--TeV mass range with $R=3$ or larger. However, as $R$ approaches the resonance region with $R \approx 2$, regions of dark photon mass do appear where it is possible to satisfy all of the constraints associated with electroweak precision observables, thermal relic density, and direct detection.  

When the dark photon couples to Dirac fermion dark matter the allowed regions are quite small and restricted to smaller values of $\alpha_D$, as illustrated in Fig.~\ref{fig:Dirac-fermion}. For scalar dark matter one also needs to be near the resonance region but the acceptable regions of dark  photon mass are considerably broader. It is worth noting that the mass region of the dark photon around 2--4 GeV, suggested in a recent global analysis of deep inelastic scattering data~\cite{Hunt-Smith:2023sdz}, is allowed in the near resonance region, with considerable flexibility in the scalar dark matter case. On the other hand, the recent LZ5T limits~\cite{LZ:2024zvo} rule out almost all scenarios with a very heavy dark photon below 1 TeV.

Our phenomenological model may be derived from a UV complete theory~\cite{Roy:2025inq}, but the details of that do not affect the phenomenological analysis presented here. In the future, the proposed experimental facilities~\cite{CEPCStudyGroup:2018ghi, FCC:2018evy} are expected to measure some of the electroweak observables with significantly increased precision, which could improve the current constraints on the dark sector. Moreover, direct detection experiments could put more emphasis on analyzing possible dark matter signals associated with effective interactions beyond the standard spin-independent (SI) and spin-dependent (SD) operators~\cite{Wang:2025cth}.
%=======================================================%

\ack{
We would like to thank Pengxuan Zhu for helpful communications.
This work was supported by the University of Adelaide and by the Australian Research Council through the Centre of Excellence for Dark Matter Particle Physics (CE200100008). B.~M.~Loizos was also supported by an Australian Government Research Training Program Scholarship.
}

%%%%%%%%%%%%%%%%%%%%%%%%%%%%%%%%%%%%%%%%%%%%%%%%%%
%%%%%%%%%%%%%%%%%%%%%%%%%%%%%%%%%%%%%%%%%%%%%%%%%%%%
\section*{References}
\bibliographystyle{iopart-num}
\bibliography{bibliography}

\end{document}